\documentclass[UTF8, a4paper]{article}  

\usepackage{geometry}
\usepackage{amsmath, amssymb, amsfonts}  
\usepackage{bm}                          
\usepackage{mathtools}                   

\usepackage{graphicx}    
\usepackage{hyperref}    
\usepackage{enumitem}    

\usepackage{multirow}
\usepackage{hhline}
\usepackage{makecell}

\usepackage{authblk}
\usepackage{hyperref}

\providecommand{\keywords}[1]{\textbf{Keywords:} #1}

\title{The gravity side of the $p$-adic AdS$_2$/CFT$_1$}
\author{Feng Qu\thanks{qufeng@syu.edu.cn}}
\affil{College of Science, Shenyang University}

\date{}

\begin{document}
	
\maketitle
	
\begin{abstract}
By means of a higher-dimensional embedding approach, together with the study of distance, measure, and curvature, we construct a gravitational theory on the two-dimensional space over $p$-adic numbers. Before constructing this gravitational theory, we provide supplementary calculations based on existing literature and justify the validity of the gravitational theory by comparison with the real-number case, for instance, the curvatures of de Sitter(dS), anti-de Sitter(AdS), and Euclidean space should be positive, negative, and zero, respectively. Finally, as an example on the gravity side of the $p$-adic AdS$_2$/CFT$_1$, we present the action for a scalar field coupled to gravity.
\end{abstract}

	
	
\tableofcontents

\section{Introduction}

The basic idea of the AdS/CFT duality~\cite{Maldacena:1997re,Gubser:1998bc,Witten:1998qj} is that there exists a correspondence between a gravitational theory in AdS space and a conformal field theory on the boundary, such that problems on each side can be transformed into those of the other. There are various versions of this duality, including the $p$-adic version $p$AdS/CFT~\cite{Gubser:2016guj,Heydeman:2016ldy}, where the discrete version of $\mathbb{Q}_p^2$ (two-dimensional space over $p$-adic numbers), namely the Bruhat-Tits tree T$_p$, is usually taken as the $p$-adic analog of AdS space in the real-number case, and the gravity side of $p$AdS/CFT is studied by investigating gravity theories on this tree~\cite{Heydeman:2016ldy,Gubser:2016htz,Huang:2019nog,Chen:2021rsy,Chen:2021qah,Chen:2021ipv}. The present work also falls into this category. However, unlike previous approaches, we first embed $\mathbb{Q}_p^2$ as the boundary into a higher-dimensional space, and then by defining distance, measure, and curvature, we construct a gravitational theory on $\mathbb{Q}_p^2$. Based on the scalar field theory proposed in~\cite{Qu:2018ned}, we present the action for a scalar field coupled to gravity.

The organization of this paper is as follows: Section~\ref{sec:embedingofqp2} discusses the embedding details of $\mathbb{Q}_p^2$; Section~\ref{sec:disonthetree} defines the distance on $\mathbb{Q}_p^2$; Section~\ref{sec:example} presents three examples of different spaces on $\mathbb{Q}_p^2$ which can be regarded as the $p$-adic version of dS, AdS and Euclidean space; in Section~\ref{sec:gravity}, after introducing measure and curvature, we construct a gravitational theory on $\mathbb{Q}_p^2$; and finally, the action for gravity coupled to a scalar field is given in Section~\ref{sec:gravityandmatter}. The last section is the summary and discussion.

\section{$\mathbb{Q}_p^2$ as the boundary of a tree}\label{sec:embedingofqp2}

\begin{figure}
	\centering
	\includegraphics[width=0.65\textwidth]{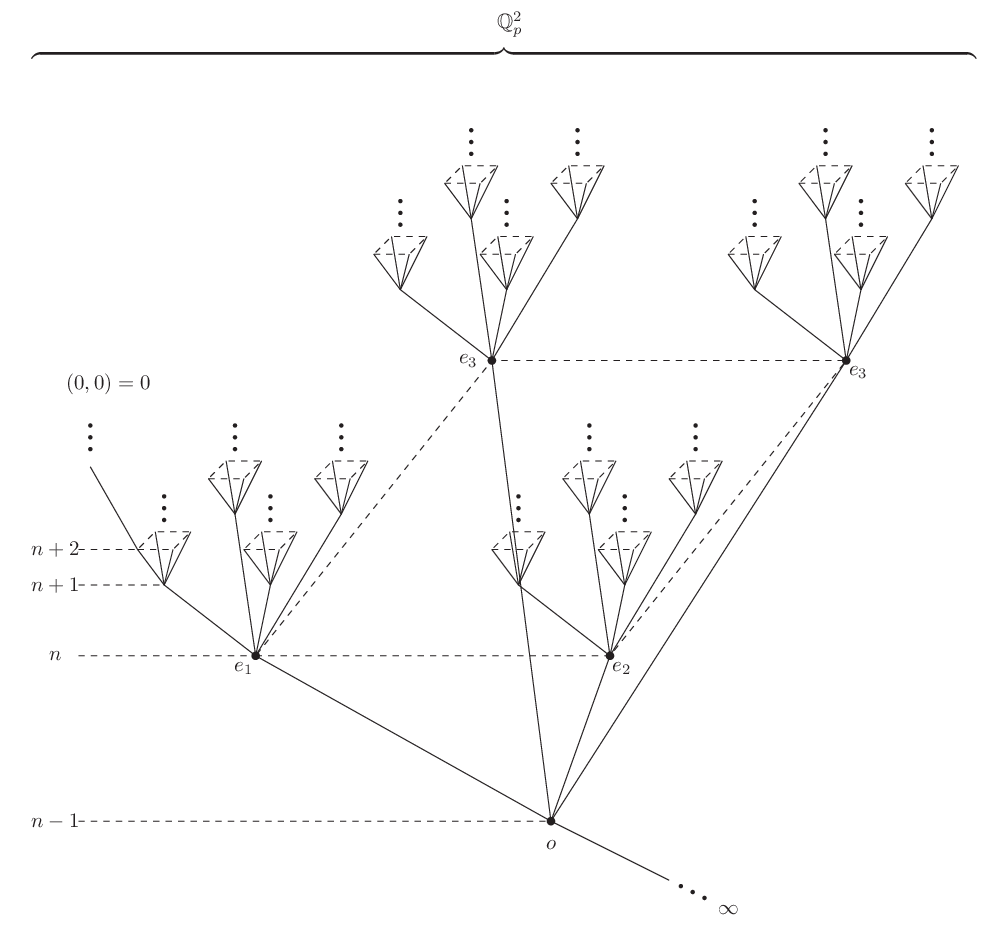}
	\caption{The tree representation of $\mathbb{Q}_p^2$ with $p=2$}
	\label{fig:tp2}
\end{figure}

Consider the two-dimensional extension of $p$-adic numbers:
\begin{gather}
	\mathbb{Q}_p^2=\{x=(x_1,x_2)\mid x_1,x_2\in\mathbb{Q}_p\}~.
\end{gather}
Just as T$_p$ is related to $\mathbb{Q}_p$, so is another tree graph related to $\mathbb{Q}_p^2$. Write $x_i$ as a $p$-adic decimal expansion:
\begin{gather}
	x_1=a_{-N}a_{-N+1}\cdots a_{-1}.a_0a_1a_2\cdots=a_{-N}p^{-N}+a_{-N+1}p^{-N+1}+\cdots+a_{0}p^{0}+\cdots~,
	\\
	x_2=b_{-M}b_{-M+1}\cdots b_{-1}.b_0b_1b_2\cdots=b_{-M}p^{-M}+b_{-M+1}p^{-M+1}+\cdots+b_{0}p^{0}+\cdots~,
	\\
	a_i,b_j=0,1,\cdots,p-1~,
	\\
	a_{-N},b_{-M}\neq0~,
	\\
	|x_1|_p=p^{N}~,~|x_2|_p=p^M~,
\end{gather}
where $a_k,b_k$ denote the digits in the $k$-th place and $|x_i|_p$ is the $p$-adic absolute value of $x_i$. Just as in the case of T$_p$, we introduce a tree graph (the Bruhat-Tits tree T$_{p^2}$) in Fig.~\ref{fig:tp2}.
If we imagine this tree as lying in three-dimensional space, the parallelograms drawn with dashed lines are actually placed horizontally. Two adjacent sides of the parallelogram lie in two directions: one parallel to the page and the other perpendicular to it. It is an infinite tree with each vertex having $1+p^2$ incident edges. The vertical coordinates ($\cdots,n-1,n,n+1,n+2,\cdots$ in the figure) of vertices take integer values, corresponding to the places in the decimal expansion. Every straight line (or polyline) from bottom to top corresponds to an element in $\mathbb{Q}_p^2$. Taking $p=2$ as an example, starting from a vertex at level $n$ (a vertex with vertical coordinate $n$) and moving upward, if $a_{n}=0$ the polygonal line deviates to the left in the direction parallel to the page; if $a_{n}=1$, it deviates to the right in the direction parallel to the page; if $b_{n}=0$, it simultaneously deviates outward in the direction perpendicular to the page; and if $b_{n}=1$, it deviates inward in the direction perpendicular to the page. Eventually, the polyline reaches one of the four adjacent vertices at level $n+1$. The values of the remaining digits in the expansion are reflected on the polyline in the same manner, and so on. Therefore, each upward polyline corresponds to one $(x_1,x_2)$, which means the upper boundary of the tree is $\mathbb{Q}_p^2$. The construction of this tree is entirely analogous to that of T$_p$ in~\cite{Gubser:2016guj}, and we denote it by T$_{p^2}$.

\begin{figure}
	\centering
	\includegraphics[width=0.7\textwidth]{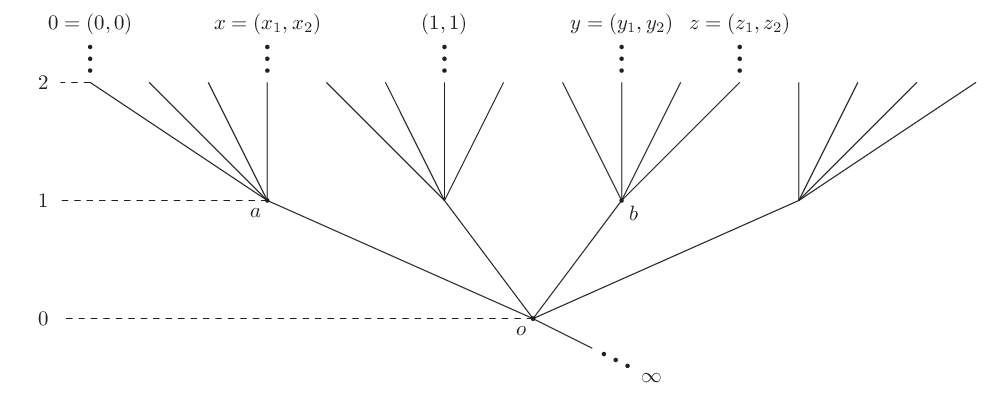}
	\caption{The relation between $|x-y|_p$ and the lowest vertex on the line connecting $x$ and $y$}
	\label{fig:tp22}
\end{figure}

Similar to the $p$-adic absolute value $|\cdot|_p$ on $\mathbb{Q}_p$, we introduce the corresponding two-dimensional version of it using the same symbol:
\begin{gather}
	|x|_p=|(x_1,x_2)|_p:=\max\{|x_1|_p,|x_2|_p\}~,~x\in\mathbb{Q}_{p}^2~.\label{eqn:abs}
\end{gather}
The symbol ``$:=$'' stands for ``is defined as''. It is not difficult to find that $|x-y|_p$ depends on the vertical coordinate of the lowest vertex on the line connecting $x$ and $y$. Letting $h(a)$ give the vertical coordinate of vertex $a$ and referring to T$_{p^2}$ in Fig.~\ref{fig:tp22},
we can write down some equations as follows:
\begin{gather}
	|x-y|_p=|(x_1,x_2)-(y_1,y_2)|_p=|(x_1-y_1,x_2-y_2)|_p=p^{-h(o)}=1~,
	\\
	|x-0|_p=|(x_1,x_2)-(0,0)|_p=|(x_1,x_2)|_p=p^{-h(a)}=p^{-1}=|y-z|_p~,
	\\
	|x-(1,1)|_p=|x-y|_p=|x-z|_p=1~.
\end{gather}
It is the same as that in the case of $\mathbb{Q}_p$ and T$_p$ tree.

In this paper, all $p$-adic numbers and their $p$-adic absolute values are dimensionless.

\section{Distance on the boundary}\label{sec:disonthetree}

\begin{figure}
	\centering
	\includegraphics[width=0.7\textwidth]{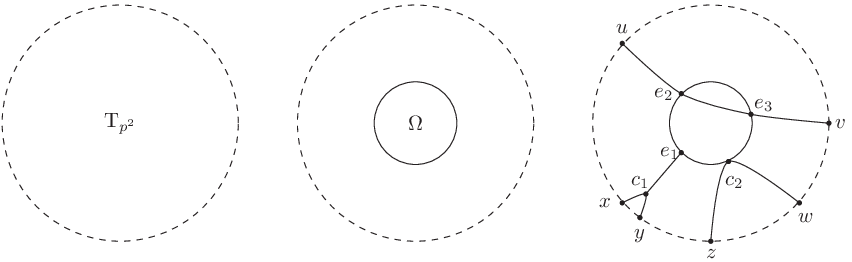}
	\caption{The subgraph-dependent regularized distance on the boundary}
	\label{fig:bi}
\end{figure}

On the tree T$_{p^2}$, let $d(A,B)$ denote the length of the path connecting $A$ and $B$, where $A$ and $B$ can be either a vertex or a boundary point. Since it diverges when $A$ and $B$ are boundary points, a regularization scheme is needed if we want to study the distance function on the boundary. We introduce a scheme depending on a subgraph of the tree. In the left panel of Fig.~\ref{fig:bi},
a dashed circle represents the boundary of the tree, and the inside of the circle is the tree itself. In the middle panel of Fig.~\ref{fig:bi}, the circle with a solid line and its interior represent a subgraph $\Omega$ of the tree. The distance between two boundary points can be regularized by the distance between $\Omega$ and the boundary points. Inspired by~\cite{Gubser:2016guj}, the regularized distance $D$ of two boundary points $x$ and $y$ is defined as
\begin{gather}
	D^2(x,y):=L^2p^{\frac{d(x,y)-d'(x,\Omega)-d'(y,\Omega)}{L}}~,\label{eqn:summaryanddiscussiondis1}
	\\
	d'(x,\Omega):=\min\{d(x,\omega)\mid\omega\in\Omega\}~.\label{eqn:summaryanddiscussiondis2}
\end{gather}
$L$ is a parameter with the dimension of length, for example, the length of a particular edge on the tree. $\omega$ runs over all vertices of $\Omega$, and $d'(x,\Omega)$ has the meaning of ``the distance between $x$ and $\Omega$''. For convenience, we will sometimes drop the prime and write $d(x,\Omega)$ when no confusion arises.

In order to demonstrate that $D$ is finite, we examine all three possible positional configurations of the line connecting two boundary points with respect to $\Omega$. As shown in the right panel of Fig.~\ref{fig:bi}, let $(A-B)$ denote the line (path) connecting $A$ and $B$. Then in the case of $(x-y)$ we have
\begin{gather}
	D^2(x,y)=L^2p^{\frac{d(x,y)-d(x,e_1)-d(y,e_1)}{L}}=L^2p^{-2\frac{d(c_1,e_1)}{L}}~,
\end{gather}
which is finite. In the case of $(z-w)$ we have
\begin{gather}
	D^2(z,w)=L^2p^{\frac{d(z,w)-d(z,c_2)-d(w,c_2)}{L}}=L^2p^0=L^2~.
\end{gather}
In the case of $(u-v)$ we have
\begin{gather}
	D^2(u,v)=L^2p^{\frac{d(u,v)-d(u,e_2)-d(v,e_3)}{L}}=L^2p^{\frac{d(e_2,e_3)}{L}}~,
\end{gather}
which is finite too. 
	
\section{Examples of spaces on $\mathbb{Q}_p^2$}\label{sec:example}

In this section we consider the simple case where all edge lengths on the tree are equal to $L$. Let $n(A,B)$ be the number of edges between the two points. Then we have
\begin{gather}
n(A,B)=\frac{d(A,B)}{L}~.
\end{gather}
The distance $D$ in this section can be written as
\begin{gather}
	D^2(x,y)=L^2p^{n(x,y)-n'(x,\Omega)-n'(y,\Omega)}~,
	\\
	n'(x,\Omega):=\min\{n(x,\omega)\mid\omega\in\Omega\}~.
\end{gather}
For convenience, we set $L=1$ in this section and, when no confusion arises, omit the prime and write $n(x,\Omega)$ in stead of $n'(x,\Omega)$.

\subsection{One vertex as the subgraph}\label{sec:vertextp2}

\begin{figure}
	\centering
	\includegraphics[width=0.8\textwidth]{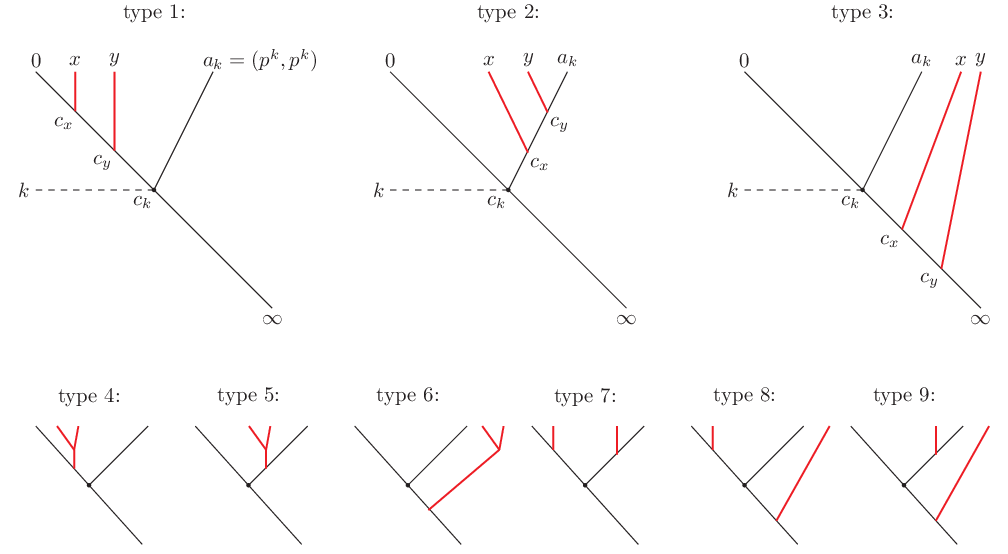}
	\caption{Different positional relations between $\Omega=\{c_k\}$ and line $(x-y)$}
	\label{fig:cases}
\end{figure}

On T$_{p^2}$, consider the case where there is only one vertex in $\Omega$. Without loss of generality, assume that it is the common vertex of the lines $((0,0)-\infty),((0,0)-(p^k,p^k))$, and $(\infty-(p^k,p^k))$ with $k\in\mathbb{Z}$, and denote it by $c_k$, namely $\Omega=\{c_k\}$. As shown in Fig.~\ref{fig:cases}, there are nine types of positional relations between $\Omega$ and $(x-y)$.
In the case of type 1, we can write down following equations:
\begin{gather}
\log_pD^2(x,y)=n(x,y)-n(x,\Omega)-n(y,\Omega)=n(x,y)-n(x,c_k)-n(y,c_k)=-2n(c_y,c_k)~,
\\
|x-y|_p=p^{-h(c_y)}~,
\\
n(c_y,c_k)=h(c_y)-h(c_k)=h(c_y)-k~.
\end{gather}
A straightforward calculation gives
\begin{gather}
D^2(x,y)=p^{2k}|x-y|_p^2~.\label{eqn:type1}
\end{gather}
By following analogous procedures, we can obtain results of all types as
\begin{gather}\label{eqn:9type}
D^2(x,y)=\left\{
\begin{aligned}
p^{2k}|x-y|_p^2~,&~\textrm{type 1,2,4,5}~,
\\
p^{-2k}\min\{|x|_p,|y|_p\}^{-2}~,&~\textrm{type 3}~,
\\
p^{-2k}|x-y|_p^2\max\{|x|_p,|y|_p\}^{-4}~,&~\textrm{type 6}~,
\\
1~,&~\textrm{type 7,8,9}~,
\end{aligned}
\right.
\end{gather}

Somewhat surprisingly, in the case of $k=0$ this piecewise function can be written in a unified form:
\begin{gather}
D^2(x,y)=\frac{|x-y|_p^2}{\max\{1,|x|_p\}^2\max\{1,|y|_p\}^2}~,~x,y\in\mathbb{Q}_p^2~.\label{eqn:disds2}
\end{gather}
For example, in the case of type 3, we have
\begin{gather}
\left.
\begin{aligned}
\max\{1,|x|_p\}=&|x|_p
\\
\max\{1,|y|_p\}=&|y|_p=|x-y|_p
\\
\min\{|x|_p,|y|_p\}=&|x|_p
\\
D^2(x,y)=&\frac{|x-y|_p^2}{\max\{1,|x|_p\}^2\max\{1,|y|_p\}^2}
\end{aligned}
\right\}\Rightarrow D^2(x,y)=\min\{|x|_p,|y|_p\}^{-2}~.
\end{gather}
The expression of $D$ in~(\ref{eqn:disds2}) is similar to the chordal distance for the Euclideanized dS$_2$ space over real numbers. 

\subsection{A boundary point  as the subgraph}\label{sec:disbdy2}

In the above calculations, if we let the vertex $c_k$ tend to the boundary point $\infty$ (or $k\to-\infty$), namely $\Omega=\{\infty\}$, using the expression for $D$ in type 1 and type 4~(\ref{eqn:9type}) which are the only two possible types when $k$ is sufficiently negative, we have
\begin{gather}
	D^2(x,y)=p^{2k}|x-y|_p^2\to0\times|x-y|_p^2~.
\end{gather}
Performing an additional regularization and removing the factor that tends to zero, we can obtain a finite distance $D'$:
\begin{gather}
(D'(x,y))^2:=p^{-2k}D^2(x,y)=|x-y|_p^2~,~x,y\in\mathbb{Q}_p^2~.\label{eqn:dism2}
\end{gather} 
The expression for $D'$ is similar to the (chordal) distance for the two-dimensional Euclidean space over real numbers.

\subsection{T$_p$ tree as the subgraph}\label{sec:tptp'}

\begin{figure}
	\centering
	\includegraphics[width=0.45\textwidth]{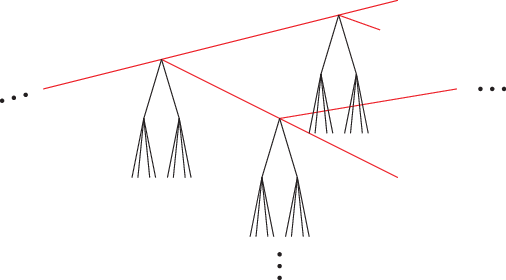}
	\caption{T$_p$ (drawn in red) as the subgraph $\Omega$ of T$_{p^2}$ with $p=2$}
	\label{fig:tp12}
\end{figure}

As shown in Fig.~\ref{fig:tp12}, this is a more complicated case. 
Fortunately, it has already been computed in \cite{Gubser:2016guj}, and the chordal distance $u_p(x,y)$ in that paper is precisely the $D^2(x,y)$ in this paper. We write its result as
\begin{gather}
D^2(x,y)=D^2((x_1,x_2),(y_1,y_2))=\frac{\max\{|x_1-y_1|_p,|x_2-y_2|_p\}^2}{|x_1y_1|_p}~.\label{eqn:disads2}
\end{gather} 
It is similar to the square of the chordal distance for the Euclideanized AdS$_2$ space over real numbers.

\subsection{Comparison with the real-number case}\label{sec:compandr}

Let us summarize the above examples and introduce some notations for these spaces over $p$-adic numbers. The results are summarized in Table~\ref{tab:sump}. 
\begin{table}[htbp]
	\centering
	\caption{\label{tab:sump}Different spaces on $\mathbb{Q}_p^2$}
	\begin{tabular}{|c|c|c|c|c|}
		\hline
		Space&Embedding&Subgraph $\Omega$&$D^2$ or $(D')^2$ depending on $\Omega$&Notation\\
		\hhline{|=|=|=|=|=|}   
		\multirow{3}{*}{$\mathbb{Q}_p^2$}&\multirow{3}{*}{$\mathbb{Q}_p^2=\partial\textrm{T}_{p^2}$} &\makecell{Vertex\\$c((0,0),(1,1),\infty)$}&$\frac{\max\{|x_1-y_1|_p^2,|x_2-y_2|_p^2\}}{\max\{1,|x_1|_p^2,|x_2|_p^2\}\max\{1,|y_1|_p^2,|y_2|_p^2\}}$&$p$dS$_2$\\
		\cline{3-5}         
		&  &\makecell{Boundary point\\$\infty$}&$\max\{|x_1-y_1|_p^2,|x_2-y_2|_p^2\}$&$p$E$_2$\\
		\cline{3-5} 
		&  &\makecell{Tree\\T$_p$}&$\frac{\max\{|x_1-y_1|_p^2,|x_2-y_2|_p^2\}}{|x_1y_1|_p}$&$p$AdS$_2$\\
		\hline
	\end{tabular}
\end{table}
$\partial\textrm{T}_{p^2}$ represents the boundary of T$_{p^2}$, and $\mathbb{Q}_p^2=\partial\textrm{T}_{p^2}$ means $\mathbb{Q}_p^2$ is embedded into T$_{p^2}$ as its boundary. $c((0,0),(1,1),\infty)$ denotes the common vertex on the lines $((0,0)-(1,1)),((0,0)-\infty)$, and $((1,1)-\infty)$. When $\Omega$ tends to the boundary point $\infty$, $D$ needs to be regularized to $D'$.

For real-number cases, we denote two-dimensional Euclidean space by E$_2$, and denote the Euclideanized two-dimensional (anti-)de Sitter space by E(A)dS$_2$. See Table~\ref{tab:sumr} for spaces on $\mathbb{R}^2$.
\begin{table}[htbp]
	\centering
	\caption{\label{tab:sumr}Different spaces on $\mathbb{R}^2$}
	\begin{tabular}{|c|c|}
		\hline
		Space notation&Square of the choral distance\\   
		\hhline{|=|=|}
		EdS$_2$&$4R^4\frac{(x_1-y_1)^2+(x_2-y_2)^2}{(R^2+x_1^2+x_2^2)(R^2+y_1^2+y_2^2)}$\\
		\hline
		E$_2$&$(x_1-y_1)^2+(x_2-y_2)^2$\\
		\hline
		EAdS$_2$&$R^2\frac{(x_1-y_1)^2+(x_2-y_2)^2}{x_1y_1}~,~x_1,y_1>0$\\
		\hline
	\end{tabular}
\end{table}
The formula in the last line applies only to half of EAdS$_2$. $R$ is the radius of (A)dS. 

By comparing Table~\ref{tab:sump} with Table~\ref{tab:sumr}, we find that, in $p$-adic cases,  different choices of the subgraph $\Omega$ lead to the boundary becoming different spaces: the $p$-adic version of dS, AdS, or Euclidean space. This can be summarized as ``Different $\Omega$'s give rise to different spaces on the boundary''.

Examples in this section impose restrictions on the notion of curvature in the next section: the curvatures on AdS, dS, and Euclidean spaces should be negative, positive, and zero, respectively.

\section{Gravitational theory on $\mathbb{Q}_p^2$}\label{sec:gravity}

In order to study the gravitational theory on the boundary $\mathbb{Q}_{p}^2$, we need to allow all edge lengths on T$_{p^2}$ to vary. However, in this paper we only consider a relatively simple case where the edge lengths on $\Omega$ are variable, while those outside $\Omega$ are fixed to $L$.

\begin{figure}
	\centering
	\includegraphics[width=0.4\textwidth]{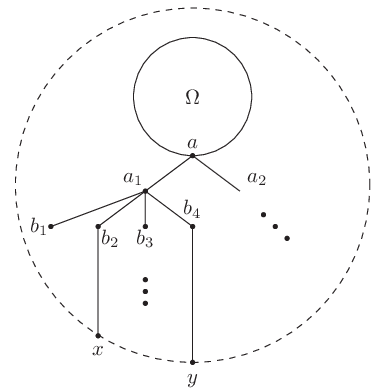}
	\caption{Vertices as balls in $\mathbb{Q}_p^2$}
	\label{fig:measure}
\end{figure}

As shown in Fig.~\ref{fig:measure}, 
$\mathbb{Q}_p^2$ is embedded into T$_{p^2}$ as the boundary, and $\Omega$ is the subgraph determining the distance on the boundary.
In the figure, $a$ is a vertex of $\Omega$, and on paths from $a$ to the boundary, some vertices are labeled as $a_i$ and $b_i$. Each vertex can be regarded as a ball on the boundary, consisting of those boundary points that are reachable from the vertex. For example we have
\begin{gather}
x\in b_2\subset a_1\subset a~,
\\
b_1\cup b_2\cup b_3\cup b_4=a_1~,
\\
a_1\cup a_2=a~.
\end{gather}

When edge lengths outside $\Omega$ are equal to a constant $L$, the tree-like structure growing from $a$ toward the boundary gives rise to the non-Archimedean nature of $a$, which means every boundary point belonging to $a$ can be regarded as the center of the ball $a$. Therefore, all boundary points belonging to $a$ are mutually equivalent, and all vertices (balls) contained in the ball $a$ and at the same distance (measured by the number of edges) from the vertex $a$ should have the same measure. Referring to Fig.~\ref{fig:measure}, we have
\begin{gather}
	\mu(a_1)+\mu(a_2)=\mu(a)~,~\mu(a_1)=\mu(a_2)~,
	\\
	\sum_i\mu(b_i)=\mu(a_1)~,~\mu(b_1)=\mu(b_2)=\mu(b_3)=\mu(b_4)~.
\end{gather}
A similar measure has been used in~\cite{Gubser:2016guj} and a subsequent paper~\cite{Qu:2018ned}. 

We can define a measure (or volume element) $dx$ on the boundary that is compatible with $\mu$. We require that
\begin{gather}
\int_{c}dx=\mu(c)~,~c\textrm{~(as a ball)~}\subset\mathbb{Q}_p^2~.
\end{gather}

\begin{figure}
	\centering
	\includegraphics[width=0.4\textwidth]{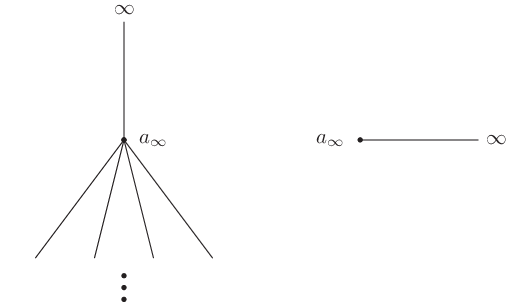}
	\caption{\label{fig:pe1}Details near the boundary and the modified subgraph.}
\end{figure}

As for the curvature on the boundary $\mathbb{Q}_p^2$, let us first deal with that on $\Omega$. Based on Table~\ref{tab:sump} and the fact that the curvatures of AdS, dS, and Euclidean space are negative, positive, and zero, respectively, we attempt to introduce a curvature on vertices of $\Omega$ as
\begin{gather}
	R(a):=1-deg(a)~,~a\in\Omega~,\label{eqn:sanddR1}
\end{gather}
where $deg(a)$ is the number of edges incident to $a$ on $\Omega$. When $\Omega=\textrm{T}_p$ and $\Omega$ contains only one vertex, we find
\begin{gather}
	p\textrm{AdS}_2:~R(a)=1-(1+p)=-p<0~,\label{eqn:padscura}
	\\
	p\textrm{dS}_2:~R(a)=1-0=1>0~.\label{eqn:pdscura}
\end{gather}
The situation for $p$E$_2$ when $\Omega$ contains only one boundary point $\infty$ is a bit more subtle. Since $\infty$ is not a tree vertex, the domain of $R$ is empty; we therefore slightly enlarge $\Omega$ to include a vertex. Taking $p=2$ as an example, as shown in Fig.~\ref{fig:pe1}, 
the left panel shows the boundary point $\infty$ and its neighboring vertex $a_{\infty}$, while the right panel shows the enlarged $\Omega$. Such a modified subgraph yields the desired curvature as
\begin{gather}
	p\textrm{E}_2:~R(a)=1-1=0~.\label{eqn:pecura}
\end{gather}

If we regard the curvature and the edge lengths on $\Omega$ as independent of each other and allow variations of $deg(a)$, then the gravitational theory would permit topological changes of $\Omega$. However, in this paper we do not consider such a complicated situation; instead, we require that the curvature also depends on the edge lengths on $\Omega$. This definition of curvature should be compatible with~(\ref{eqn:sanddR1}). The choice is not unique. Let $b\sim a$ denote that $b$ is a neighboring vertex of $a$ on $\Omega$. We consider a relatively simple case in which we redefine the curvature as
\begin{gather}
R(a):=1-\sum_{\substack{b\sim a\\b\in\Omega}}\frac{d(a,b)}{L}~,~a\in\Omega~,\label{eqn:summaryanddiscussionr1}
\end{gather}
where~(\ref{eqn:sanddR1}) can be recovered as the special case of $d(a,b)=L$. Note that $d(a,b)$ is the edge length between $a$ and $b$ when $b\sim a$. 

Next, we define $R$ on the boundary. As shown in Fig.~\ref{fig:measure}, considering that all boundary points belonging to $a$ are equivalent, the curvature at these points should also be identical. We define the curvature on boundary points as
\begin{gather}
	R(x):=R(a)~,~x\in a~.\label{eqn:sanddR2}
\end{gather}

By analogy with the theory of gravity over the real numbers, where $S\sim\int dx^n\sqrt{-g}R$, we can write down the theory of gravity on $\mathbb{Q}_p^2$ as
\begin{gather}
	S\sim\int dxR(x)~.
\end{gather}
Given a subgraph $\Omega$, this action on the boundary can be reformulated as an action on $\Omega$. Ignoring the arbitrary constant, we have
\begin{gather}
	S=\int dxR(x)=\sum_{a\in\Omega}\int_adxR(x)=\sum_{a\in\Omega}\mu(a)R(a)~.
\end{gather}
There are many different choices for $\mu(a)$. We consider the simple choice
\begin{gather}
	\mu(a)=\sum_{\substack{b\sim a\\b\in\Omega}}d(a,b)~,~a\in\Omega~.\label{eqn:summaryanddiscussionmeasure}
\end{gather}
Then the action can be written as
\begin{gather}
	S=\sum_{a\in\Omega}\sum_{\substack{b\sim a\\b\in\Omega}}d(a,b)\Big(1-\sum_{\substack{c\sim a\\c\in\Omega}}\frac{d(a,c)}{L}\Big)~.\label{eqn:sanddaction}
\end{gather}
This is a gravitational theory both on the boundary $\mathbb{Q}_p^2$ and on the subgraph $\Omega$. 

\section{Coupling to the scalar field}\label{sec:gravityandmatter}

A scalar field theory on $p$AdS$_2$ has been proposed in~\cite{Qu:2018ned}, where all edge lengths on T$_{p^2}$ are equal and fixed. Using the volume element $dx$ and the distance $D$ introduced in this paper, we generalize the result of~\cite{Qu:2018ned} to the case where the edge lengths of $\Omega$ are variable. The action of a scalar field can be written as
\begin{gather}
S_{\textrm{matter}}=c\int dx\int_{y\in F(x)} dy\frac{(\phi(x)-\phi(y))^2}{D^{2\alpha}(x,y)}+\int dx\frac{1}{2}m^2\phi^2(x)~,
\end{gather}
where $c$ is a constant and $\alpha$ is a parameter related to the derivative order. The integration region $y\in F(x)$ requires clarification. Denote the vertex on $\Omega$ containing $x$ by $a$, and denote its $deg(a)$ neighboring vertices on $\Omega$ by $b_{1},b_{2},\cdots$. The integration region is the union of them, which can be written as
\begin{gather}
F(x)=a\cup b_{1}\cup b_{2}\cdots\cup b_{deg(a)}~.
\end{gather}
The integral form of the $\frac{\phi(x)-\phi(y)}{D(x,y)}$ term is related to the properties of the space. Inside the ball $a$, the space is non-Archimedean; small‑distance propagations cannot be combined to yield large-distance propagation. Hence, the propagation inside $a$ must include the differences of the two fields, namely $\phi(x)-\phi(y)$, at all distances. This is the reason why $a$ is contained in $F(x)$. On $\Omega$, if the field can propagate from the ball 
$a$ to its neighboring vertex (ball) $b_{i}$, then it can propagate to other vertices on $\Omega$ in the same way, that is, to arbitrarily far distances. This is the reason why $b_{i}$'s are contained in $F(x)$ and also the reason why $F(x)$ does not need to be extended further. Considering that
\begin{gather}
\int_{y\in F(x)}dy\cdots=\int_{y\in a}dy\cdots+\sum_{\substack{b\sim a\\b\in\Omega}}\int_{y\in b}dy\cdots~,
\end{gather} 
we need to consider the expression for $D(x,y)$ in different cases. 

In the case of $y\in a$ (corresponding to the case of $(x-y)$ in the right panel of Fig.~\ref{fig:bi}), we have
\begin{gather}
D^2(x,y)=L^2p^{\frac{d(x,y)-d(x,\Omega)-d(y,\Omega)}{L}}=L^2p^{-2n(c_1,e_1)}~,
\end{gather}
which is independent of the edge length on $\Omega$. Note that $n(c_1,e_1)$ is the number of edges between $c_1$ and $e_1$ in the right panel of Fig.~\ref{fig:bi}. In the case of $y\in b_{i}$ (corresponding to the case of $(u-v)$ in the right panel of Fig.~\ref{fig:bi}), we have
\begin{gather}
	D^2(x,y)=L^2p^{\frac{d(x,y)-d(x,\Omega)-d(y,\Omega)}{L}}=L^2p^{\frac{d(a,b_{i})}{L}}~,
\end{gather}
where $d(a,b_{i})$ is the length of the edge connecting $a$ and $b_{i}$. Considering $\int dx\cdots=\sum_{a\in\Omega}\int_{x\in a}dx\cdots$, the matter action can be written as
\begin{gather}
S_{\textrm{matter}}=c\sum_{a\in\Omega}\int_{x\in a}dx\int_{y\in a} dy\frac{(\phi(x)-\phi(y))^2}{L^{2\alpha}p^{-2\alpha n(c_1,e_1)}}+c\sum_{a\in\Omega}\int_{x\in a}dx\sum_{\substack{b\sim a\\b\in\Omega}}\int_{y\in b}dy\frac{(\phi(x)-\phi(y))^2}{L^{2\alpha}p^{\alpha\frac{d(a,b)}{L}}}+\int dx\frac{1}{2}m^2\phi^2(x)~.
\end{gather}
Absorbing $1/L^{2\alpha}$ into $c$, and adding $S_{\textrm{matter}}$ to the action~(\ref{eqn:sanddaction}), we obtain the action for the gravity coupled to a scalar field as
\begin{gather}
\begin{aligned}
S_{\textrm{tot}}=&\sum_{a\in\Omega}\sum_{\substack{b\sim a\\b\in\Omega}}d(a,b)\Big(1-\sum_{\substack{c\sim a\\c\in\Omega}}\frac{d(a,c)}{L}\Big)
\\
+&c\sum_{a\in\Omega}\int_{x\in a}dx\int_{y\in a} dy\frac{(\phi(x)-\phi(y))^2}{p^{-2\alpha n(c_1,e_1)}}+c\sum_{a\in\Omega}\int_{x\in a}dx\sum_{\substack{b\sim a\\b\in\Omega}}\int_{y\in b}dy\frac{(\phi(x)-\phi(y))^2}{p^{\alpha\frac{d(a,b)}{L}}}+\int dx\frac{1}{2}m^2\phi^2(x)~.
\end{aligned}\label{eqn:tot}
\end{gather}
As shown in the right panel of Fig.~\ref{fig:bi}, $n(c_1,e_1)$ is the number of edges along the shortest path connecting $\Omega$ and the line $(x-y)$.

\section{Summary and discussion}

In this paper, we have mainly accomplished the following tasks:

First, we embed $\mathbb{Q}_p^2$ into the tree T$_{p^2}$ as its boundary. With the help of a subgraph $\Omega$, we define the distance $D$ on the boundary, as shown in~(\ref{eqn:summaryanddiscussiondis1}) and~(\ref{eqn:summaryanddiscussiondis2}). 

Second, when all edge lengths are equal, three examples of different spaces on $\mathbb{Q}_p^2$ are given in Table~\ref{tab:sump}, while the corresponding examples on $\mathbb{R}^2$ are given in Table~\ref{tab:sumr}.

Third, when the edge lengths on $\Omega$ are variable and those outside $\Omega$ are fixed to $L$, we define the curvature $R$ in~(\ref{eqn:summaryanddiscussionr1}) and (\ref{eqn:sanddR2}). For the $p$-adic versions of AdS, dS, and Euclidean spaces, it gives the correct signs: negative, positive, and zero, respectively. After introducing the measure in~(\ref{eqn:summaryanddiscussionmeasure}), the gravitational action on $\mathbb{Q}_p^2$ is obtained in~(\ref{eqn:sanddaction}).

Finally, based on the scalar field theory on $p$AdS$_2$ in~\cite{Qu:2018ned}, we write down the action for gravity coupled to a scalar field in~(\ref{eqn:tot}).

There are at least two further issues that remain to be discussed. One is to allow the edge lengths outside $\Omega$ to vary and to introduce more complicated curvature and measure; the other is to perform some detailed calculations in the context of $p$‑adic AdS/CFT and compare the results with those in the real‑number case.

\section*{Acknowledgments}

This work was completed with the help of DeepSeek-R1-0528, which provided several candidate actions on tree graphs for the author to consider. The final selection and all conclusions are the sole responsibility of the author.



\begin{thebibliography}{99}
	
\bibitem{Maldacena:1997re}
J.~M.~Maldacena,
``The Large $N$ limit of superconformal field theories and supergravity,''
Adv. Theor. Math. Phys. \textbf{2} (1998), 231-252
doi:10.4310/ATMP.1998.v2.n2.a1
[arXiv:hep-th/9711200 [hep-th]].

\bibitem{Gubser:1998bc}
S.~S.~Gubser, I.~R.~Klebanov and A.~M.~Polyakov,
``Gauge theory correlators from noncritical string theory,''
Phys. Lett. B \textbf{428} (1998), 105-114
doi:10.1016/S0370-2693(98)00377-3
[arXiv:hep-th/9802109 [hep-th]].

\bibitem{Witten:1998qj}
E.~Witten,
``Anti de Sitter space and holography,''
Adv. Theor. Math. Phys. \textbf{2} (1998), 253-291
doi:10.4310/ATMP.1998.v2.n2.a2
[arXiv:hep-th/9802150 [hep-th]].
	
	
\bibitem{Gubser:2016guj}
S.~S.~Gubser, J.~Knaute, S.~Parikh, A.~Samberg and P.~Witaszczyk,
``$p$-adic AdS/CFT,''
Commun. Math. Phys. \textbf{352} (2017) no.3, 1019-1059
doi:10.1007/s00220-016-2813-6
[arXiv:1605.01061 [hep-th]].



\bibitem{Heydeman:2016ldy}
M.~Heydeman, M.~Marcolli, I.~Saberi and B.~Stoica,
``Tensor networks, $p$-adic fields, and algebraic curves: arithmetic and the AdS$_3$/CFT$_2$ correspondence,''
Adv. Theor. Math. Phys. \textbf{22} (2018), 93-176
doi:10.4310/ATMP.2018.v22.n1.a4
[arXiv:1605.07639 [hep-th]].

\bibitem{Gubser:2016htz}
S.~S.~Gubser, M.~Heydeman, C.~Jepsen, M.~Marcolli, S.~Parikh, I.~Saberi, B.~Stoica and B.~Trundy,
``Edge length dynamics on graphs with applications to $p$-adic AdS/CFT,''
JHEP \textbf{06} (2017), 157
doi:10.1007/JHEP06(2017)157
[arXiv:1612.09580 [hep-th]].

\bibitem{Huang:2019nog}
A.~Huang, B.~Stoica and S.~T.~Yau,
``General relativity from $p$-adic strings,''
Adv. Theor. Math. Phys. \textbf{26} (2022) no.5, 1203-1237
doi:10.4310/ATMP.2022.v26.n5.a4
[arXiv:1901.02013 [hep-th]].

\bibitem{Chen:2021rsy}
L.~Chen, X.~Liu and L.~Y.~Hung,
``Bending the Bruhat-Tits tree. Part I. Tensor network and emergent Einstein equations,''
JHEP \textbf{06} (2021), 094
doi:10.1007/JHEP06(2021)094
[arXiv:2102.12023 [hep-th]].

\bibitem{Chen:2021qah}
L.~Chen, X.~Liu and L.~Y.~Hung,
``Bending the Bruhat-Tits tree. Part II. The p-adic BTZ black hole and local diffeomorphism on the Bruhat-Tits tree,''
JHEP \textbf{09} (2021), 097
doi:10.1007/JHEP09(2021)097
[arXiv:2102.12024 [hep-th]].

\bibitem{Chen:2021ipv}
L.~Chen, X.~Liu and L.~Y.~Hung,
``Emergent Einstein Equation in p-adic Conformal Field Theory Tensor Networks,''
Phys. Rev. Lett. \textbf{127} (2021) no.22, 221602
doi:10.1103/PhysRevLett.127.221602
[arXiv:2102.12022 [hep-th]].

\bibitem{Qu:2018ned}
F.~Qu and Y.~h.~Gao,
``Scalar fields on $p$AdS,''
Phys. Lett. B \textbf{786} (2018), 165-170
doi:10.1016/j.physletb.2018.09.043
[arXiv:1806.07035 [hep-th]].

\end{thebibliography}
\end{document}